# Anisotropy and oblique total transmission at a planar negative-index interface

Lei Zhou, C.T. Chan and P. Sheng

Department of Physics, The Hong Kong University of Science and Technology
Clear Water Bay, Kowloon, Hong Kong, China

**Abstract**

We show that a class of negative index ($n$) materials has interesting anisotropic optical properties, manifest in the effective refraction index that can be positive, negative, or purely imaginary under different incidence conditions. With dispersion taken into account, reflection at a planar negative-index interface exhibits frequency selective total oblique transmission that is distinct from the Brewster effect. Finite-difference-time-domain simulation of realistic negative-$n$ structures confirms the analytic results based on effective indices.

Materials whose permittivity $\varepsilon$ and permeability $\mu$ are simultaneously negative are said to possess a negative refractive index $n$, with many unusual properties [1]. Negative-*n* metallic resonating composites and two dimensional (2D) isotropic negative-*n* material have been constructed [2,3], and negative light refraction was observed [4]. The unconventional properties of such materials have drawn an increasing amount of attention in both science and engineering [5-10].

Recently, the anisotropic wave characteristics of 2D isotropic negative-*n* materials [3-4] have been studied [8,9]. A bilayer structure composed of anisotropic negative-*n* materials has been proposed to be a perfect lens [10]. In this work, we are interested in the optical anisotropy and reflection/refraction properties at a planar interface with a negative-*n* material, such as those described in Ref. [2], whose wave characteristics have yet to be examined. We find the material to possess an angle-dependent effective refractive index $n(\theta)$, which can be either negative, positive, or purely imaginary at different wave-vector directions. While it was pointed out that most negative-*n* interfaces had reflectance near unity for incident propagating wave [10], we find the present one to exhibit frequency selective oblique total transmission, whose governing physics is different from that of the Brewster angle phenomenon [11]. Finite-difference-time-domain (FDTD) simulations [12] on a realistic metallic resonance structure confirmed the predicted effect.

Consider a homogeneous media characterized by a diagonal permittivity matrix with $\varepsilon_{yy} < 0, \varepsilon_{xx} = \varepsilon_{zz} = 1$, and a diagonal permeability matrix with $\mu_{xx} < 0, \mu_{yy} = \mu_{zz} = 1$ [13]. For EM waves traveling along the *z* direction with *E* field polarized along the *y*-axis, the material exhibits simultaneous negative $\varepsilon$ and $\mu$, leading to a negative *n*. This corresponds exactly to the metallic meta-material recently fabricated [2]. The wave equation for *B* field can be written as:

$$-\vec{k} \times [\varepsilon^{-1}(\vec{k} \times (\mu^{-1}\vec{B}))] = \omega^2 \vec{B} / c^2, \qquad (1)$$

where *c* is the speed of light, $\varepsilon^{-1}, \mu^{-1}$ are respectively the inverses of permittivity and permeability matrices, $\omega$ and $\vec{k}$ are angular frequency and wave vectors. Let $\vec{k}$ be confined in the $xz$ plane and $\overrightarrow{E}$ polarized along the $y$ axis. This will be the case studied in this paper. We note that $\vec{E} \parallel \hat{y}$ implies $B_y = 0$, and $\nabla \cdot \vec{B} = 0$ means $k_x B_x = -k_z B_z$. Putting the above two expressions into Eq. (1), we get $[(\varepsilon_{yy}\mu_{xx})^{-1}k_z^2 + (\varepsilon_{yy}\mu_{zz})^{-1}k_x^2]\vec{B} = \omega^2 \vec{B}/c^2$, which leads to a dispersion relation $\omega^2 = (ck)^2/n(\theta)^2$ with an angle-dependent refraction index

$$n(\theta)^2 = [(\varepsilon_{yy}\mu_{xx})^{-1}\cos^2\theta + (\varepsilon_{yy}\mu_{zz})^{-1}\sin^2\theta]^{-1}. \qquad (2)$$

Here, $\theta$ is the angle between $\vec{k}$ and the $z$ axis. Figure 1 shows a comparison of this angle dependent effective refraction index with those for air and for an ordinary non-absorbing anisotropic material, where the angle represents the wave-vector direction and radius gives the modulus of the index. For the ordinary non-absorbing anisotropic material, $n(\theta)$ is real in the entire angle regime, and resembles an ellipsoid [11]. Anisotropic negative-$n$ material, however, has a real $n(\theta)$ (its sign will be identified later) in some angular range, implying the support for traveling waves, and has a purely imaginary $n(\theta)$ in others, forbidding any traveling waves. In addition, if $|\varepsilon_{yy} \bullet \mu_{xx}| < 1$, we find that the condition $|n(\theta)| = 1$ can be met at some particular angles. The implication of this condition is addressed later.

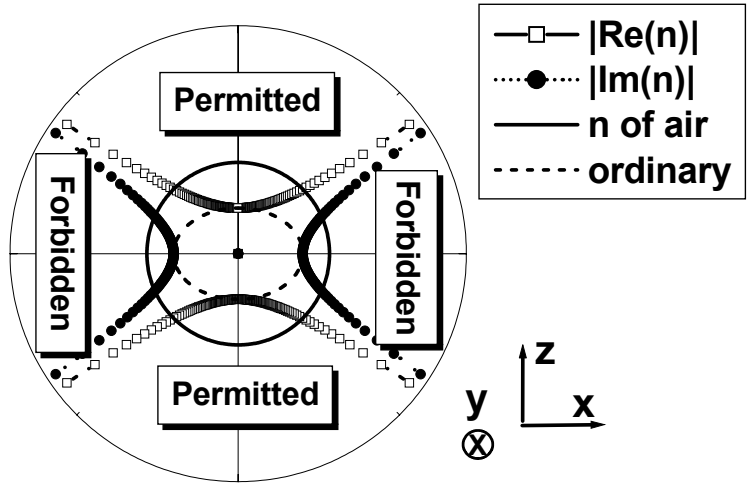

**Fig. 1 Refractive index as a function of the wave-vector direction for a negative-*n* material with $\varepsilon_{yy} = \mu_{xx} = -0.5, \mu_{zz} = 1$ (open squares for the real part, solid circles for the imaginary part). For comparison, the refractive index as a function of angle is also shown for air (solid line), and an ordinary anisotropic material with $\varepsilon_{yy} = \mu_{xx} = 0.5, \mu_{zz} = 1$ (dashed line).**

We now consider the reflection/refraction properties at a planar interface between air and such a negative-*n* material. Due to the special angular variation in *n*, there are two distinct cases corresponding to two different orientations of the interface. Consider first a planar *xy* interface between air and a negative-*n* material. Translational invariance along surface directions at the interface means the conservation of the parallel $\vec{k}$ component:

$$\theta_i = \theta_r, |\sin(\theta_i)| = |n(\theta_t^k)\sin(\theta_t^k)|, \qquad (3)$$

where $\theta_i, \theta_r$ and $\theta_t^k$ are respectively the incident, reflection and refraction angles. The ray diagram is schematically depicted in Fig. 2, where the subscripts *i,r,t* stand for the incident, reflected and transmitted waves, respectively. With $k_t^x$ fixed by $k_i^x$, there are two choices for the sign of $k_t^z$ in the second media. Causality requires that the Poynting vector $\vec{S} \equiv \vec{E} \times \vec{H}$ (different from the $\vec{k}$ vector!) inside the second media should point away from the interface so that the transmitted wave carries energy away from the source. Based on this criterion, we choose $k_t^z > 0$ for this case. It is rather easy to check that the other choice ($k_t^z < 0$) gives an unphysical direction for $\vec{S}_t$ (subscript denotes transmission). We note that a negative $\varepsilon_{yy}$ means $\vec{D}_t$ is antiparallel to $\vec{E}_t$ for the transmitted wave; and $\mu_{xx} \neq \mu_{zz}$ indicates that $\vec{B}_t$ is non-collinear with $\vec{H}_t$. The Poynting vector is specified by an angle, $\theta_t^s$, which is distinct from $\theta_t^k$ for the wavevector and should be carefully determined [14]. It should be noted that if $\theta_t^s$ were regarded as the refraction angle, then the refraction here corresponds to a positive index.

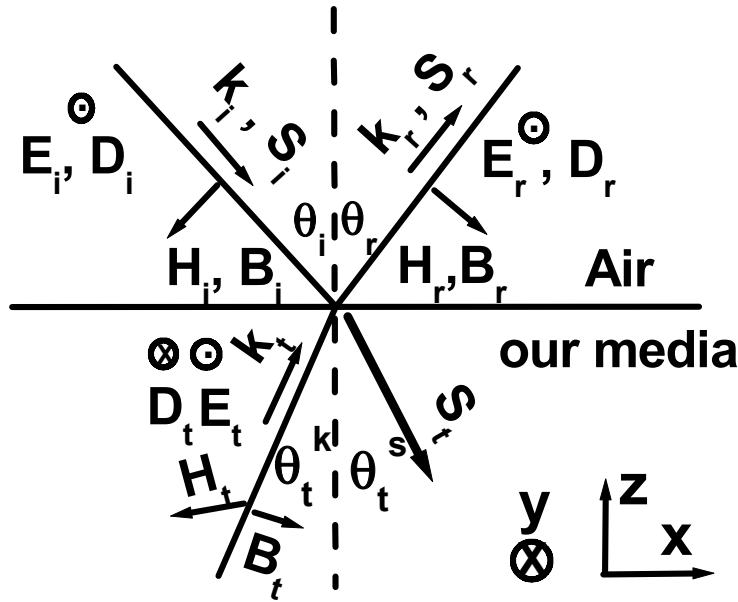

**Fig. 2 Schematic ray diagram for light reflection/refraction at the *xy* interface between air and the present studied negative-*n* material. The interface is marked by the solid line and the surface normal by the dotted line.**

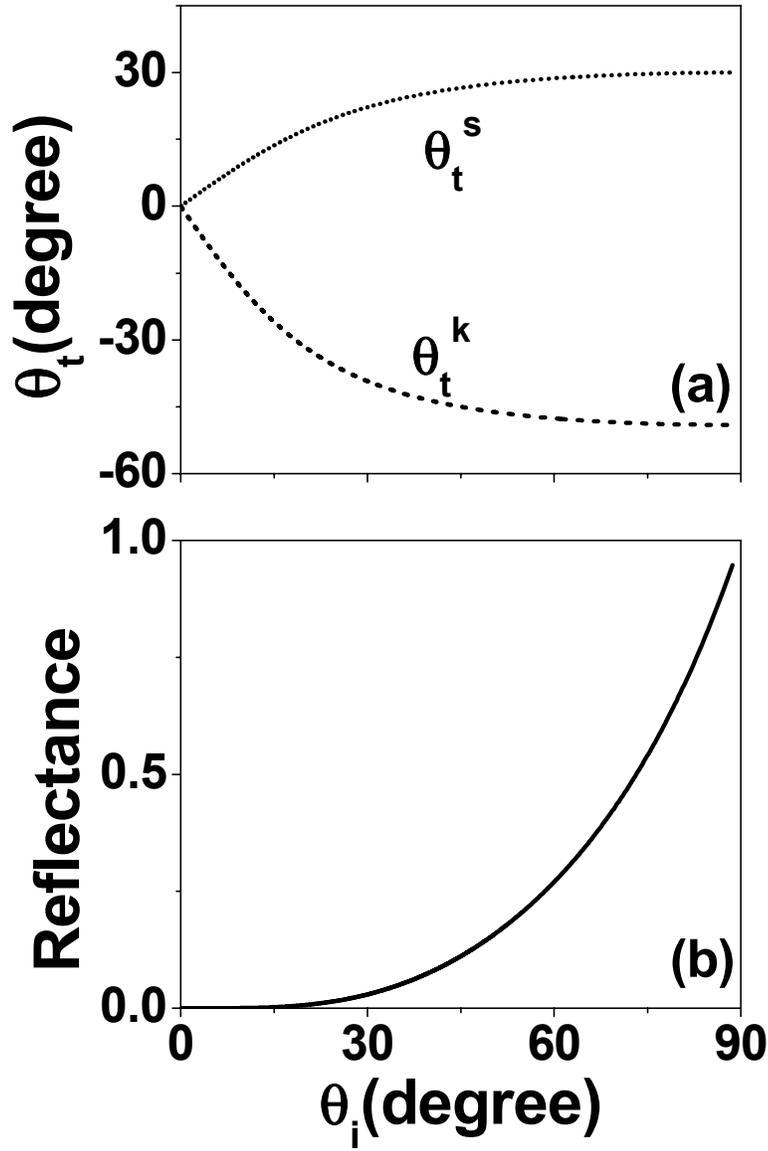

**Fig. 3 (a) Refraction angle $\theta_t^s$ (dotted line), $\theta_t^k$ (dashed line) and (b) the reflectance as functions of the incidence angle $\theta_i$ at the *xy* interface between air and a negative-*n* material with $\varepsilon_{yy} = \mu_{xx} = -0.5, \mu_{zz} = 1$.**

Applying the usual boundary conditions to the parallel components of *E* and *H* at the interfaces, the Fresnel formula for the reflectance *R* is derived to be

$$R = (\frac{|n(\theta_t^k)| \cos \theta_t^k - \cos \theta_i |\mu_{xx}|}{|n(\theta_t^k)| \cos \theta_t^k + \cos \theta_i |\mu_{xx}|})^2 \ . \qquad (4)$$

For the reflectance shown in Fig. 3(b), we note that $R$ increases as $\theta_i$ increases, similar to that of an ordinary material, and the refracted beams are confined inside the permitted regime [see Fig. 3(a)]. Since $|n(\theta_t^k)|$ is an increasing function of $\theta_t^k$ and diverges when $\theta_t^k$ approaches the boundary of the permitted regime, it is clear that one can always find a $\theta_t^k$ for any $\theta_i$ which satisfies Eq. (3).

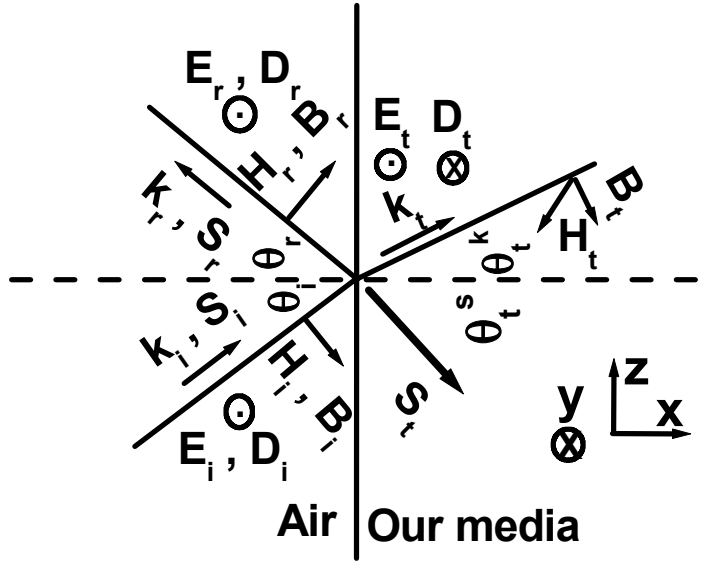

**Fig. 4 Schematic ray diagram for light reflection/refraction at the *yz* interface between air and the negative-*n* material.**

Let us now consider the second case in which the interface is at the *yz* plane. The ray diagram is shown schematically in Fig. 4, where $\theta_t^k$ is related to $\theta_i$ via

$$|\sin \theta_i| = |n'(\theta_t^k) \sin \theta_t^k| , \qquad (5)$$

in which $n'(\theta)^2 = [(\varepsilon_{yy}\mu_{zz})^{-1} \cos^2 \theta + (\varepsilon_{yy}\mu_{xx})^{-1} \sin^2 \theta]^{-1}$, in which the angles are consistently defined with respect to the normal to the interface. Again, the choice $k_t^x > 0$ adopted here is determined by the condition that $\vec{S}_t$ should point away from the interface in the second media. The most striking feature in this case is the negative refraction (see the direction of $\vec{S}_t$ in Fig. 4) [15]. A simple calculation gives the corresponding Fresnel formula to be

$$R = (\frac{|n^{'}(\theta_t^k)| \cos\theta_t^k - \cos\theta_i |\mu_{zz}|}{|n^{'}(\theta_t^k)| \cos\theta_t^k + \cos\theta_i |\mu_{zz}|})^2 . \qquad (6)$$

Reflection at this interface (Fig. 5(b)) shows an unusual incidence angle dependence in the case of $|\varepsilon_{yy} \bullet \mu_{xx}| < 1$. That is, while the interface totally reflects the EM waves incident from the normal, it can be shown that there exists a critical value for the incidence angle, $\theta_i^c = \sin^{-1}[\sqrt{\varepsilon_{yy}\mu_{xx}}]$, above which Eq. (5) can be satisfied, implying that a refracted waves can be coupled into the negative-$n$ medium (see Fig. 5(a)). Distinct from the conventional case, here $\theta_t^k$ is a decreasing function of $\theta_i$, caused by the unusual angular dependence of $n^{'}(\theta_t^k)$. The fact that there exists a particular incident angle $\theta_i^{0}$ such that $n^{'}(\theta_i^{0}) = 1$ indicates that a solution $\theta_t^k = \theta_i = \theta_i^{0}$ can be found for Eq. (5). According to Eq. (6), the material becomes non-reflecting at this incidence angle (remembering that $\mu_{zz} = 1$). Away from this particular incidence angle, $\theta_t^k$ deviates quickly from $\theta_i$, leading to strong reflection.

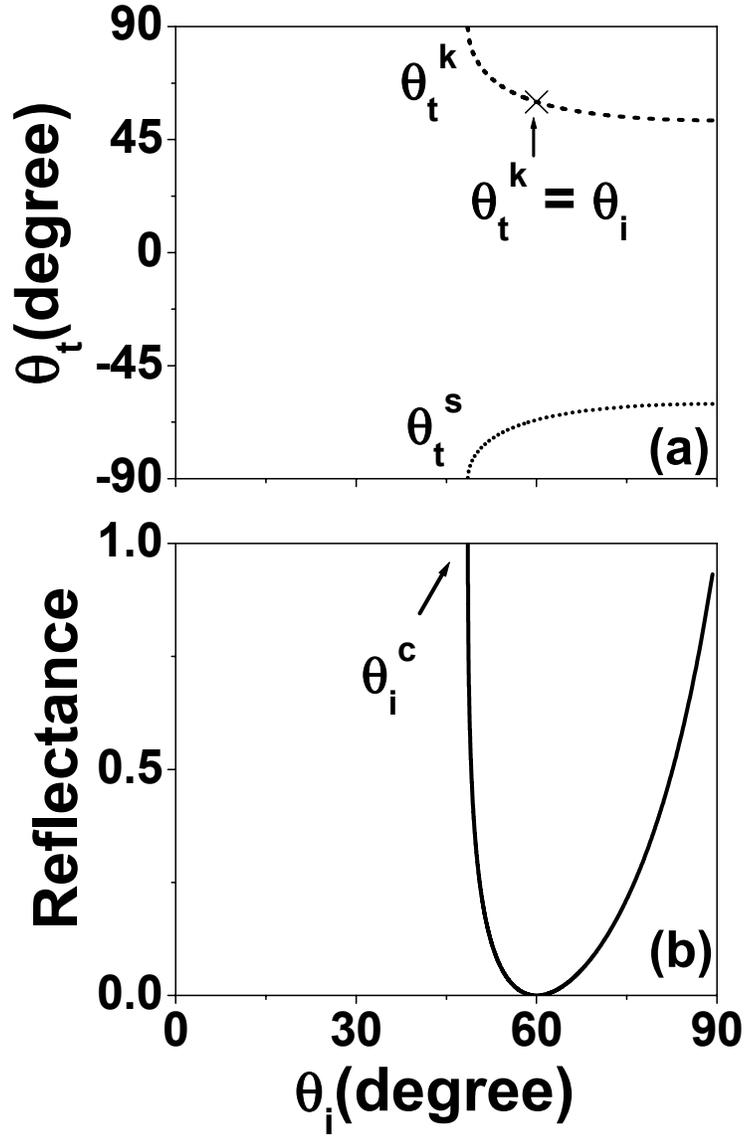

**Fig. 5 (a) Refraction angle $\theta_t^s$ (dotted line), $\theta_t^k$ (dashed line) and (b) the reflectance as functions of the incidence angle $\theta_i$ at the $yz$ interface between air and a negative-n material with $\varepsilon_{yy} = \mu_{xx} = -0.75, \mu_{zz} = 1$. Perfect transmission occurs at $\theta_i = \theta_t^k$.**

We emphasize that the physics underlying the absence of reflection here is intrinsically an anisotropic effect and is different from that for the zero reflectivity at the Brewster angle in a conventional isotropic material [11]. At the interface between air and a conventional isotropic material characterized by $\varepsilon$ and $\mu$, the reflectance for the S-polarized wave ($\vec{E} \parallel \hat{y}$) and the P-polarized wave ($\vec{H} \parallel \hat{y}$) are respectively:

$$R_S = (\frac{\cos\theta_i - \cos\theta_t / Z_t}{\cos\theta_i + \cos\theta_t / Z_t})^2, R_P = (\frac{\cos\theta_i - \cos\theta_t \cdot Z_t}{\cos\theta_i + \cos\theta_t \cdot Z_t})^2, \quad (7)$$

where $\sin\theta_i = n_t \sin\theta_t$ and $n_t = \sqrt{\varepsilon} \cdot \sqrt{\mu}, Z_t = \sqrt{\mu} / \sqrt{\varepsilon}$ are respectively the refraction index and the impedance. Zero-reflection takes place at an incidence angle satisfying $\theta_i + \theta_t = \pi / 2$ for the P wave incident on a conventional dielectric material with $\mu = 1$. A simple generalization shows that the same phenomenon exists for the S wave incident on a magnetic material (with $\varepsilon = 1$). Such an angle, determined by $\theta_B = \tan^{-1}(n_t)$, is called the Brewster angle [11]. The physics accounting for such a phenomenon is that the vibration of electrons in the second media can not generate the reflected beam which travels perpendicular to the transmitted beam (because of $\theta_i + \theta_t = \pi / 2$) [11]. The zero-reflection discovered here, however, requires $\theta_i = \theta_t$. The physics here is governed by the anisotropy which makes the material transparent at a particular oblique incidence angle (dictated by the condition $n'(\theta_t) = 1$) and dark at others (see Fig. 1). The present effect can not exist in an isotropic material. If one seeks a zero-reflection solution satisfying $\theta_i = \theta_t$ from Eq. (7) which describes an isotropic material interface, the only possibility is $\varepsilon = \mu = 1$ [16]. In fact, if one removes the anisotropy from our studied system described in Fig. 5, the total transmission phenomenon disappears (independent of whether one takes $\mu_{xx} = \mu_{zz} = 1$ or $\mu_{xx} = \mu_{zz} = -0.5$). Also, it should be noted that the present effect can only exist in a dispersive media, since the condition $|\varepsilon_{yy} \bullet \mu_{xx}| < 1$ is usually a characteristic of a dispersive media. Anisotropy in a dispersive media is the key element to get the total transmission effect.

All negative-$n$ materials are highly dispersive [3,4]. The fact that $\varepsilon$ and $\mu$ can vary from very negative values to very positive ones near the resonance suggests that any such material could always have a frequency window in which the conditions $\varepsilon_{yy}, \mu_{xx} < 0$ and $\varepsilon_{yy} \bullet \mu_{xx} < 1$ are satisfied. In what follows, we consider a 32mm-thick slab composed by a dispersive anisotropic material with effective $\varepsilon_{yy}, \mu_{xx}$ given by

$$\varepsilon_{yy}(f) = 1 + \frac{20}{3.78^2 - f^2} + \frac{100}{12^2 - f^2},$$

$$\mu_{xx}(f) = 1 + \frac{7}{3.8^2 - f^2}$$

where $f$ denotes the frequency measured in GHz. A simple calculation shows that both $\varepsilon_{yy}$ and $\mu_{xx}$ are negative and $\varepsilon_{yy} \cdot \mu_{xx} < 1$ in frequency range of 4.39—4.63 GHz. We employed the transfer matrix method to numerically calculate the transmittance through such a 32mm-thick slab as the function of the incidence angle. Within the range of 4.39—4.63 GHz, we find that there is always a specific incidence angle for which the slab becomes totally non-reflecting. Dashed line in Fig. 6(a) shows the transmittance through the slab as the function of the incidence angle for $f = 4.4$ GHz. Total transmission occurs at about $\theta_i = 75°$. Strong reflection is seen at other incidence angles [17].

Due to the dispersion of $\varepsilon$ and $\mu$, the transmission is frequency selective under a fixed incidence angle. Dashed line in Fig. 6(b) shows the transmission spectra through the 32mm-thick slab for $\theta_i = 30°$. We find total transmission at about $f = 4.55$ GHz, and almost zero transmission outside of the frequency range 4.5—4.64 GHz. We notice that the auxiliary transmission peaks in the spectra are induced by the Fabry-Perot interferences [11] between the multiply scattered fields on slab's two interfaces. These auxiliary peaks are dependent on the slab thickness, whereas the main peak is independent on the slab thickness. The solid line in Fig. 6 is an average over the transmittances of 50 slabs, each with a random thickness deviation ($\pm 20\%$). We see that the peaks due to Fabry-Perot resonances disappear upon averaging, leaving behind only the peak due to the aforementioned non-reflecting condition.

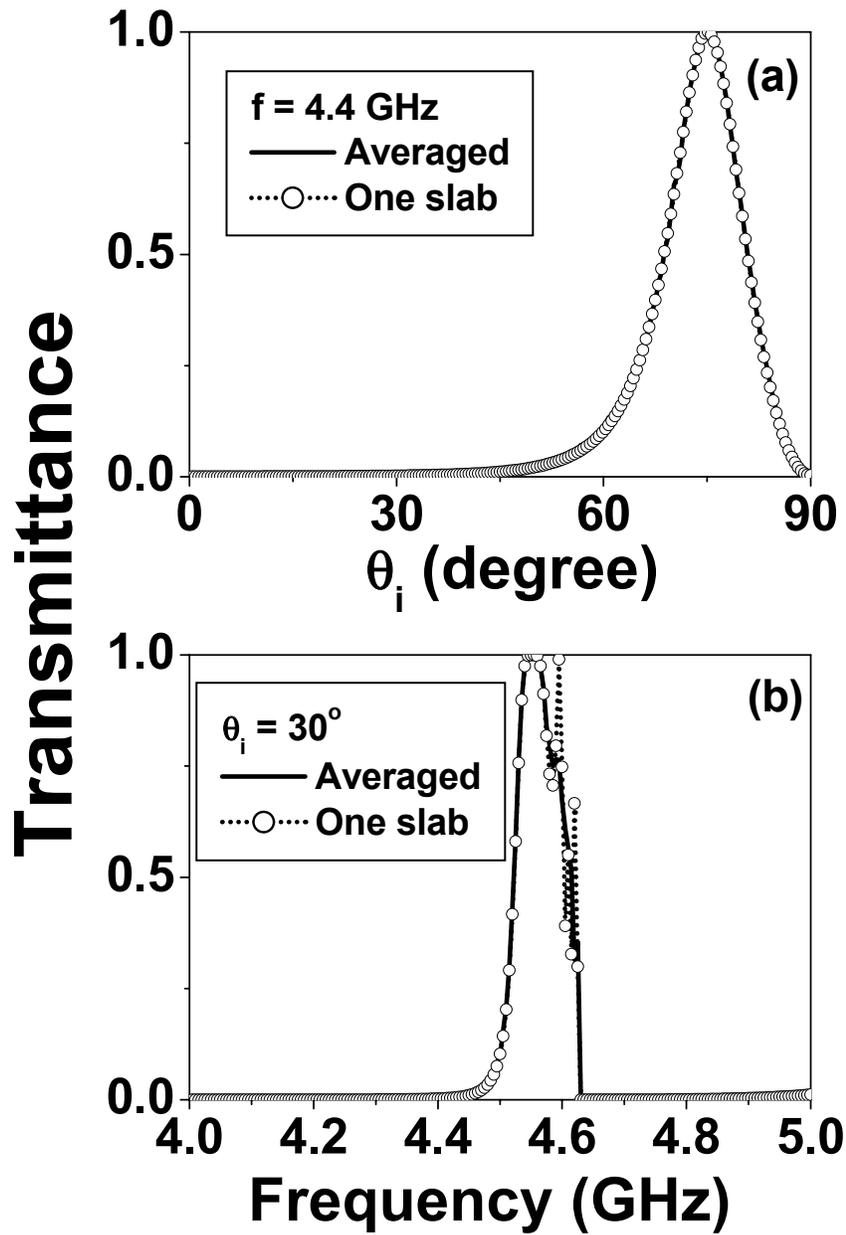

**Fig. 6** Transmittance (a) as a function of $\theta_i$ at $f = 4.4$ GHz, and (b) as a function of frequency for $\theta_i = 30°$. Dashed line is for a single 32mm-thick slab, solid line denotes the result averaged over 50 slabs with $\pm 20\%$ thickness variations.

We have performed FDTD simulations [12] to demonstrate the total transmission effect on a realistic resonance structure. The building block of our designed material is shown in the inset to Fig. 7. Here we used a metallic fork in the middle to create a negative $\varepsilon$, and split rings [18] on the left and right to create a negative $\mu$. The unit cell is then repeated with lattice constants $d_y = 16, d_z = 7.5$ mm to tile the $yz$ plane, with two 1.6mm-thick dielectric plates (with $\varepsilon = 4$) employed as substrates to separate the fork and split-rings. Finally, the resulting 3.8mm slab is repeated in the $x$ direction with lattice constant $d_x = 6$ mm to construct a layered structure. We first employed the FDTD simulation [19] to calculate the normal transmission spectra of a slab of the designed material with three unit cells along the $z$ direction (22.5mm-thick) and infinite in the $xy$ plane. From these calculated transmission spectra (using both amplitudes and phases) we then derived $\varepsilon_{yy}$ and $\mu_{xx}$ as functions of frequency. These are shown in Fig. 7. We find a frequency range where both $\varepsilon_{yy}$ and $\mu_{xx}$ are negative and $\varepsilon_{yy} \cdot \mu_{xx} < 1$. We then consider another slab of the same material with three unit cells along the $x$ direction (18mm-thick) and infinite in the $yz$ plane. The transmittance through such a slab at frequency 4.51 GHz is calculated by the FDTD simulations as a function of the incidence angle. The result is shown in Fig. 8 by the solid squares. Reasonably good agreement is seen with the result obtained for a homogeneous slab (shown by the solid line) with effective properties shown in Fig. 7. Both show the total transmission at some oblique incidence angle. It should be emphasized that the FDTD results are the numerical solutions of Maxwell's equations with the microstructures fully taken into account. The only approximation is the perfect-metal boundary condition imposed on the metal surfaces, which holds well in the microwave regime. This demonstrates that the total oblique transmission can indeed occur at a negative-$n$ interface.

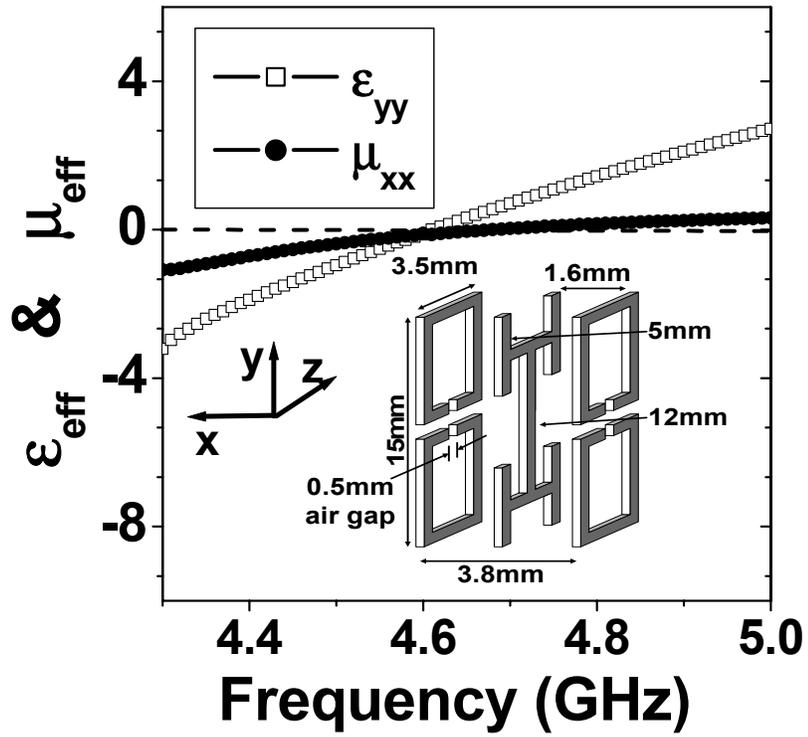

**Fig. 7** Effective $\varepsilon_{yy}$ and $\mu_{xx}$ as functions of frequency for our designed material from the FDTD simulation results. The building block structure of the designed material is shown in the inset.

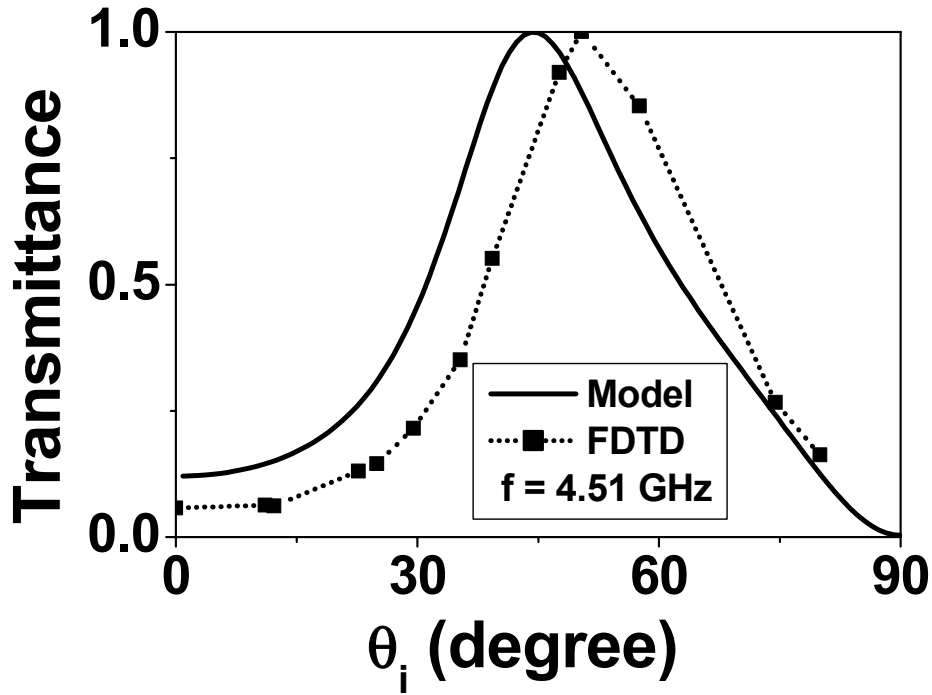

**Fig. 8 Transmittance at 4.51 GHz as a function of the incidence angle through a slab of our designed material with three unit cells in the *x* direction, calculated by the FDTD simulations (solid squares) and with effective $\varepsilon_{yy}$ and $\mu_{xx}$ shown in Fig. 7 (solid line).**

In short, we have shown that a class of anisotropic negative-*n* material has interesting angle dependent optical properties. In particular, it allows frequency selective total oblique transmission, which is different from the Brewster effect. Such a unique property has been demonstrated in a realistic metallic system with the help of FDTD simulations.

**Acknowledgement**: This work was supported by Hong Kong RGC through HKUST6145/99P and CA02/03.SC01.